\documentclass[conference]{IEEEtran}
\usepackage{times}
\usepackage{amsmath}
\usepackage{graphicx}
\usepackage{cite}
\usepackage{booktabs}
\usepackage[linesnumbered,ruled]{algorithm2e}
\usepackage{algpseudocode}
\usepackage{multirow}
\usepackage{flushend}

\begin{document}

\title{A General Scheme for Noise-Tolerant Logic Design Based on Probabilistic and DCVS Approaches}

\author{
Xinghua Yang, Fei Qiao, Qi Wei, Huazhong Yang\\
Institute of Circuit and System, Dept of Electronic Engineering, Tsinghua University\\
Tsinghua National Laboratory for Information Science and Technology\\
Email: qiaofei@tsinghua.edu.cn
}

\maketitle

\begin{abstract}
The performance of logic function could be affected significantly by the noise effect as the dimension of CMOS devices scales to nanometers. Thus, many pertinent researches about noise-tolerant logic gate have received growing attention. Considering the randomness as the noise's nature, probabilistic-based approach proves better noise-immunity and  three design schemes with the technique of Markov Random Field (MRF) have been proposed in \cite{nepal2007designing,wey2009design,liu2013general}. In this paper, a general circuit scheme for noise-tolerant logic design based on MRF theory and Differential Cascode Voltage Switch (DCVS) technique has been proposed, which is an extension of the work in \cite{liu2013general,lu2012design}. A DCVS block with only four transistors has been successfully inserted to the original circuit scheme from \cite{liu2013general} and extensive simulation results based on HSPICE show that our proposed design can operate correctly with the input signal of 1dB SNR. When using the Kullback-Leibler Distance (KLD) \cite{kullback1997information} as the evaluation parameter, the KLD value of our design decreases by $76.5\%$ on average than \cite{liu2013general} which means that superior noise-immunity could be obtained through our work.
 \end{abstract}

\section{Introduction}

Numerous performance improvements have been achieved due to the scaling down of CMOS devices in the past decades, but the problem caused by noise effect which may create fatal errors during the circuit operation become significant. Moreover, the decrease on supply voltage also deteriorates the noise-immunity of the circuit since the noise does not decrease proportionally with the supply voltage as explained in \cite{bhaduri2007reliability} and guardband voltage has to be utilized in order to keep the output correct. Thus, extensive pertinent researches about the noise-tolerant circuit design have received growing attention.

In traditional points of view, the Triple-Majority-Redundance (TMR) or Cascade TMR (CTMR) from \cite{lyons1962use,abraham1974algorithm} is a direct idea where the original computation block is duplicated triple or more times and then a vote is made based on the majority results. However, the voter could be also contaminated by the noise and this makes it inappropriate to design noise-immunity circuit. The technique of Razor proposed in \cite{blumer1987occam} has demonstrated big power reduction due to the non-conservative dynamic voltage scaling with the protective mechanism of error-rate monitoring and recovery, but this is a solution mainly focused on sequential logic level and has little effects on basic logic gate such as inverter or nand. In \cite{gupta2011impact,yang2014iscas}, approximate computing is used  to achieve more performance or energy-efficiency improvements with part of output precision losses, which can be applied to some fault tolerated designs such as multimedia, recognition or data mining processing, but the error is introduced intentionally by the circuit designers and the inherent noise could create serious erroneous output if the upper bits of the calculation results are not protected from noise impact.

Considering the randomness as the noise's nature, the aforementioned methods are hard to achieve efficient circuit immunity. Thus, some new approaches have been proposed based on probabilistic theory. The Probabilistic CMOS (PCMOS) illustrated in \cite{palem2009probabilistic} is an early attempt to exploit the random nature of the CMOS devices to obtain more design space, but the significant parts of the whole computation have to keep correct and further research of noise-immunity to the logic gate cannot be avoided in this work. Another probabilistic-based approach is proposed in \cite{nepal2007designing,wey2009design,liu2013general} where the noise-immunity of the circuit is constructed by the Markov Random Field (MRF) theory. The MRF theory is developed in \cite{li1995markov} and adopted in \cite{nepal2007designing} to solve the problems of the noise impact in logic gate design. As pointed out in \cite{nepal2007designing}, the energy of noise signal could be reduced by maximizing the joint probability of the input-output pairs, which comes at a cost of redundant hardware. This work is further optimized in \cite{wey2009design,liu2013general} where parts of the gates in the original circuit scheme \cite{nepal2007designing} are removed. In \cite{lu2012design}, the MRF approach is combined with the technique of Differential Cascode Voltage Switch (DCVS), however, this method only proposed an improved inverter to design xor-nxor gate.

In this paper, a general circuit scheme for noise-tolerant logic design based on MRF theory and DCVS technique has been proposed. A simple DCVS block with only four transistors has been successfully inserted to the Cost-Effective Noise-Tolerant Circuit based on Markov Random Field (CENT\_MRF) in \cite{liu2013general}. Extensive simulations have been implemented on HSPICE and the results show that our proposed design could operate correctly with the input signal of 1dB SNR. When using the Kullback-Leibler Distance (KLD) as the evaluation parameter, the KLD of our design decreases by $76.5\%$ on average than \cite{liu2013general} and superior noise-immunity is presented.

The remainder of the paper is organized as follows: Section 2 reviews the critical related works. Section 3 describes our proposed circuit scheme. Simulation results will be illustrated in Section 4. Finally, conclusions are drawn in section 5.

\section{Preliminary Works}

In this section, some related work in \cite{nepal2007designing,wey2009design,liu2013general,lu2012design} will be described. The original MRF theory will be illustrated first and then the process of mapping this theory into logic circuits \cite{nepal2007designing} will be explained, along with the method to design cost-effective MRF circuit structure in \cite{wey2009design,liu2013general}. At last, the work in \cite{lu2012design} will be introduced.

\subsection{MRF Theory and Corresponding Logic Circuits}

\begin{figure}[tp]
  \centering
  \includegraphics[width=3in]{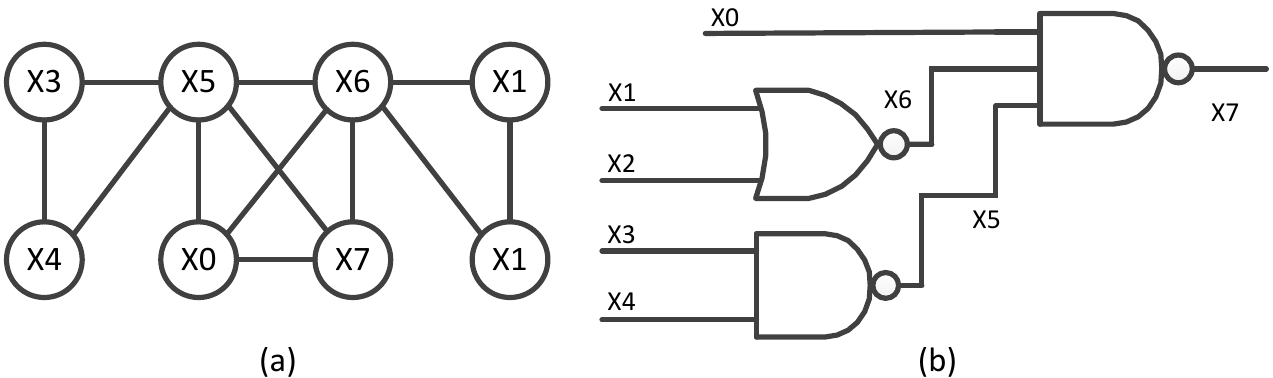}
  \caption{MRF graph and possible logic circuit. (a) an example of MRF graph; (b) apossible corresponding logic circuit.}
  \label{fig:fig1_mrf_mapping}
  \vspace*{-0.5cm}
\end{figure}

\begin{figure}[bp]
  \centering
  \vspace*{-0.5cm}
  \includegraphics[width=3in]{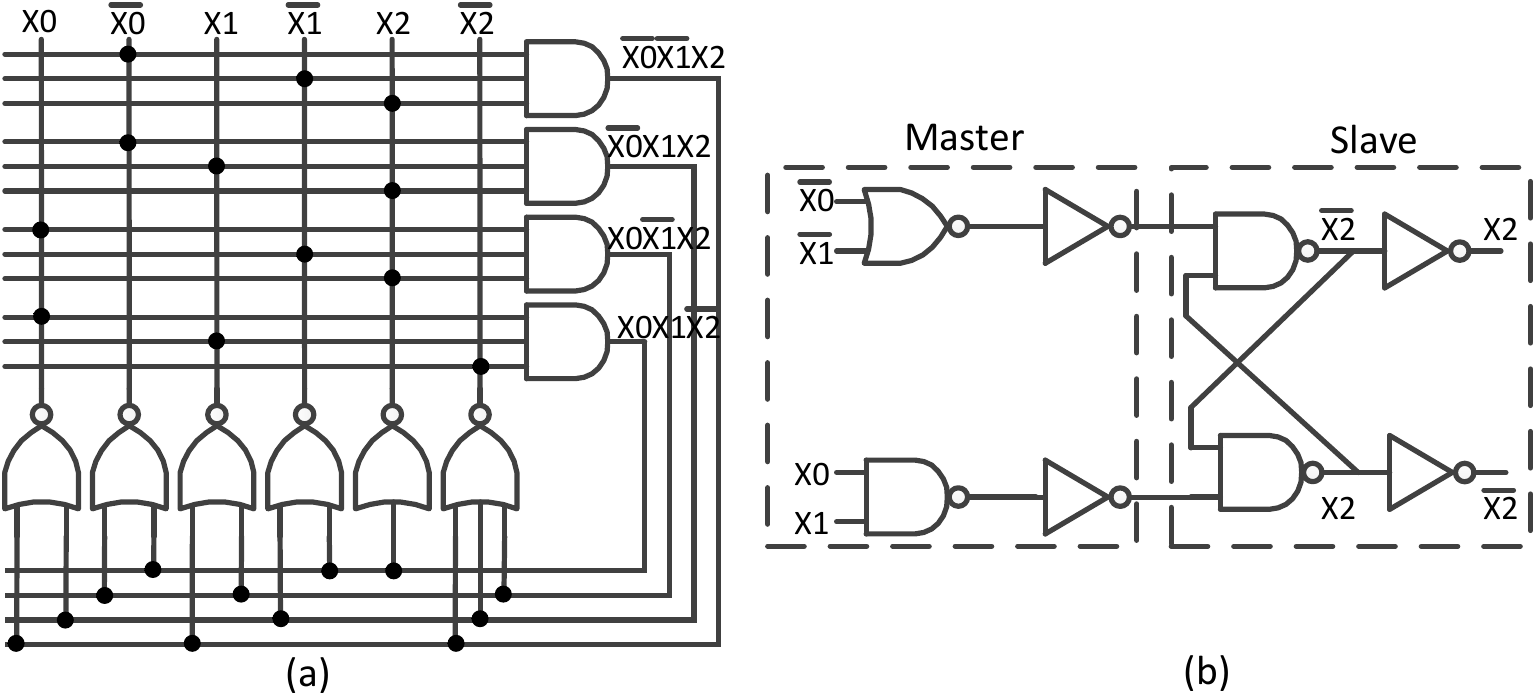}
  \caption{MRF NAND-gate. (a) MRF NAND-gate in \cite{nepal2007designing}; (b) Cost effective MRF NAND-gate in \cite{wey2009design}.}
  \label{fig:fig2_mrf_nand}
\end{figure}

Let us define a network containing a set of variables $X=\{x_0,x_1,...,x_k\}$ which are connected to each other in a certain mode as shown in Fig.\ref{fig:fig1_mrf_mapping}(a). Each variable $x_i$ can take different values from a specific set $\Omega$ (for example, in digital logic design $\Omega=\{0,1\}$) and also has its neighborhood called $N_i$ (the set of the variables connected to $x_i$). A set of variable $x_i$ and its $N_i$ is called \textit{clique}. With all of these definitions, $X$ is called a MRF if $\forall_{i}P(x_i) > 0$ and $P(x_i|\{X-x_i\}) = P(x_i|N_i)$ \cite{li1995markov}. According to the MRF theory, the joint probability of $X:P(x_1,x_2,..,x_k)$ could be maximized if every \textit{clique} has the lowest energy $U(x_c)$, which only depends all the variables in \textit{clique} $c$. As pointed in \cite{nepal2007designing}, the combinational logic circuit could be mapped to this MRF graph as shown in Fig.\ref{fig:fig1_mrf_mapping}(b). Noise-immunity can be achieved through "\textit{valid minterm feedback loop}" by which the logic gate network could be equipped with MRF property and the final correct logic state has lower energy than any other incorrect state (this means the correct output will get the highest probability). For example, in Fig.\ref{fig:fig2_mrf_nand}(a), the conventional NAND-gate is mapped onto a MRF logic network \cite{nepal2007designing} where the \textit{valid minterms} $\{\bar x_0 \bar x_1 x_2,\bar x_0 x_1 x_2,x_0 \bar x_1 x_2,x_0 x_1 \bar x_2\}$ are generated and feedback to input signals. This scheme has proved excellent noise-immunity since the final value of each node will tend to converge into the correct logic state due to its MRF property. However, this structure is inappropriate to be used in practical design as too many redundant gates are needed with this direct mapping method. Thus, in \cite{wey2009design}, this scheme is simplified and a \textit{master-and-slave} cost effective MRF design is proposed as shown in Fig.\ref{fig:fig2_mrf_nand}(b). The principles of this simplification are as followings: for the MRF NAND-gate in Fig.\ref{fig:fig2_mrf_nand}(a), the energy function of this \textit{clique} is the summation over all the \textit{valid minterms} $\{\bar x_0 \bar x_1 x_2,\bar x_0 x_1 x_2,x_0 \bar x_1 x_2,x_0 x_1 \bar x_2\}$:
\begin{equation}
\label{eqution:energy_original}
U(x_0,x_1,x_2) = -(\bar x_0 \bar x_1 x_2 + \bar x_0 x_1 x_2 + x_0 \bar x_1 x_2 + x_0 x_1 \bar x_2)
\end{equation}
After applying the Boolean difference, the Eq.\ref{eqution:energy_original} can be re-written as
\begin{equation}
\label{eqution:energy_sim_1}
U(x_0,x_1,x_2) = -((\bar x_0 + \bar x_1) x_2 + (x_0 x_1) \bar x_2)
\end{equation}
Based on Eq.\ref{eqution:energy_sim_1}, the four \textit{valid minterms} are merged into two terms $(\bar x_0+\bar x_1)$ and $(x_0 x_1)$. In Fig.\ref{fig:fig2_mrf_nand}(b), these two terms are generated by AND gate and OR gate as the Master part, while the feedback loop connected to the output of Master part is called Slave. With this methodology, large amount of hardware redundancy will be removed as the transistors can be reduced from 60 to 28. The noise-immunity of this circuit scheme is worsen than \cite{nepal2007designing} but very close as described in \cite{wey2009design}.

\subsection{CENT\_MRF in \cite{liu2013general} and MRF Circuit with DCVS in \cite{lu2012design}}

\begin{figure}[tp]
  \centering
  \includegraphics[width=3.5in]{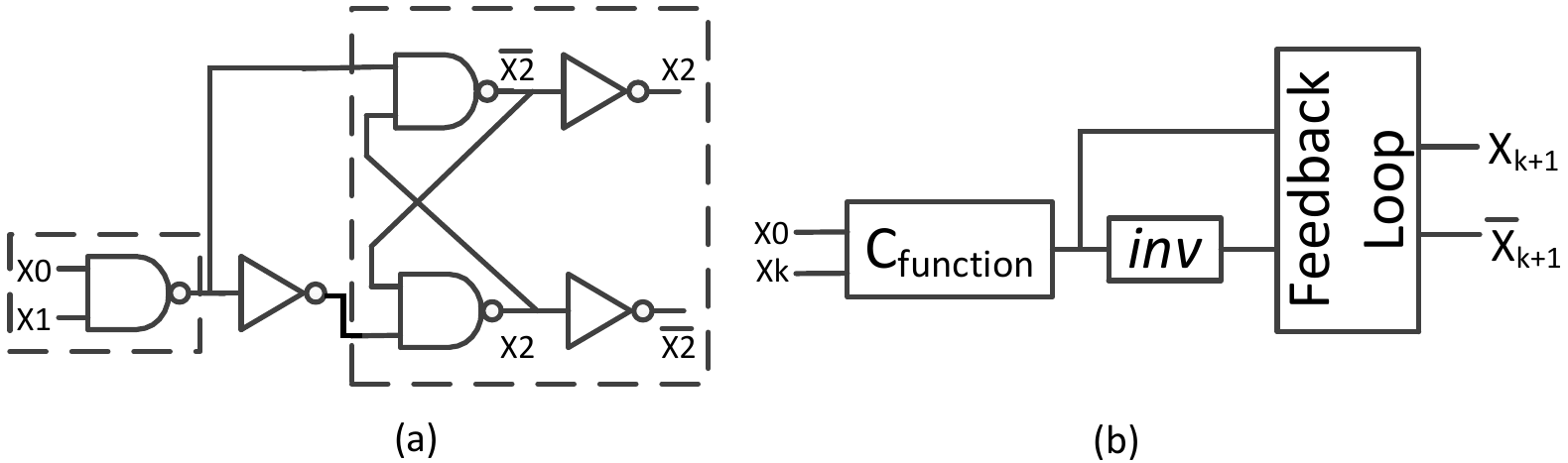}
  \caption{CENT\_MRF in \cite{liu2013general}. (a) CENT\_MRF NAND-gate; (b) general scheme for CENT\_MRF.}
  \vspace*{-0.5cm}
  \label{fig:fig3_liukaikai}
\end{figure}

In \cite{liu2013general}, the master-and-slave scheme for noise tolerant design from \cite{wey2009design} is further simplified as the Eq.\ref{eqution:energy_sim_1} can be re-written:
\begin{equation}
\begin{aligned}
\label{eqution:energy_sim_2}
U(x_0,x_1,x_2) &= -((\bar x_0 + \bar x_1) x_2 + (x_0 x_1) \bar x_2) \\
&= -(\overline {x_0 x_1} x_2 + (x_0 x_1) \bar x_2) \\
&= -(nand(x_0,x_1) x_2 + \overline {nand(x_0,x_1)}\bar x_2)
\end{aligned}
\end{equation}
Thus, the master part in Fig.\ref{fig:fig2_mrf_nand}(b) can be reconstructed with only one $nand$ gate and an $inverter$ as shown in Fig.\ref{fig:fig3_liukaikai}(a). Based on DeMorgan's Law, the energy function of basic combinational logic gates can be expressed in a general form:
\begin{equation}
\begin{aligned}
\label{eqution:energy_general}
U(x_0,x_1,...,x_k) = -(C_{function}(x_0,x_1,...,x_k)x_{k+1} \\
+\overline {C_{function}(x_0,x_1,...,x_k)}\overline {x_{k+1}})
\end{aligned}
\end{equation}
From Eq.\ref{eqution:energy_general}, a general scheme for Cost-Effective Noise-Tolerant Circuit based on MRF (CENT\_MRF) is proposed in \cite{liu2013general} as shown in Fig.\ref{fig:fig3_liukaikai}(b). Less transistors are needed in this scheme than \cite{wey2009design} but the effect of noise-immunity is worsen than \cite{wey2009design} due to the simplification.

Since the effect of noise-immunity from \cite{wey2009design,liu2013general} is worsen than the original design in \cite{nepal2007designing} although numerous transistors have been removed, some compensations for the losing immunity are made in \cite{lu2012design} by DCVS technique. However, this method only focused on inverter which is then applied to xor-nxor design and failed to build a general scheme. In next section, our proposed scheme will be described.

\section{Proposed Circuit Scheme}

In this section, a DCVS block with four transistors has been inserted to the circuit in Fig.\ref{fig:fig3_liukaikai}(a) and then a general circuit scheme based on MRF and DCVS technique will be described.

\subsection{Inserting DCVS Block into MRF-Based Circuit}
\begin{figure}[tp]
  \centering
  \includegraphics[width=1.5in]{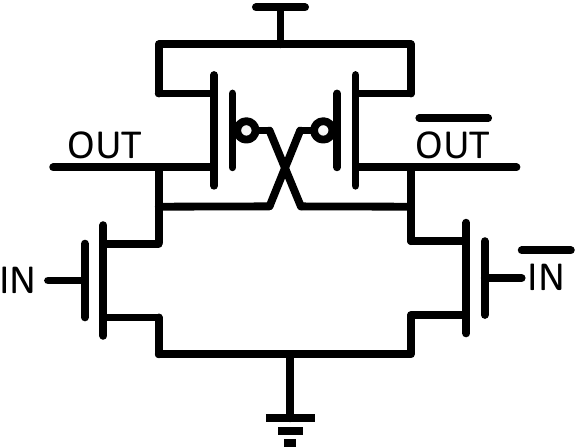}
  \caption{Differential cascode voltage switch block.}
  \vspace*{-0.5cm}
  \label{fig:fig4_dcvs}
\end{figure}

\begin{figure}[bp]
  \centering
  \vspace*{-0.5cm}
  \includegraphics[width=3.5in]{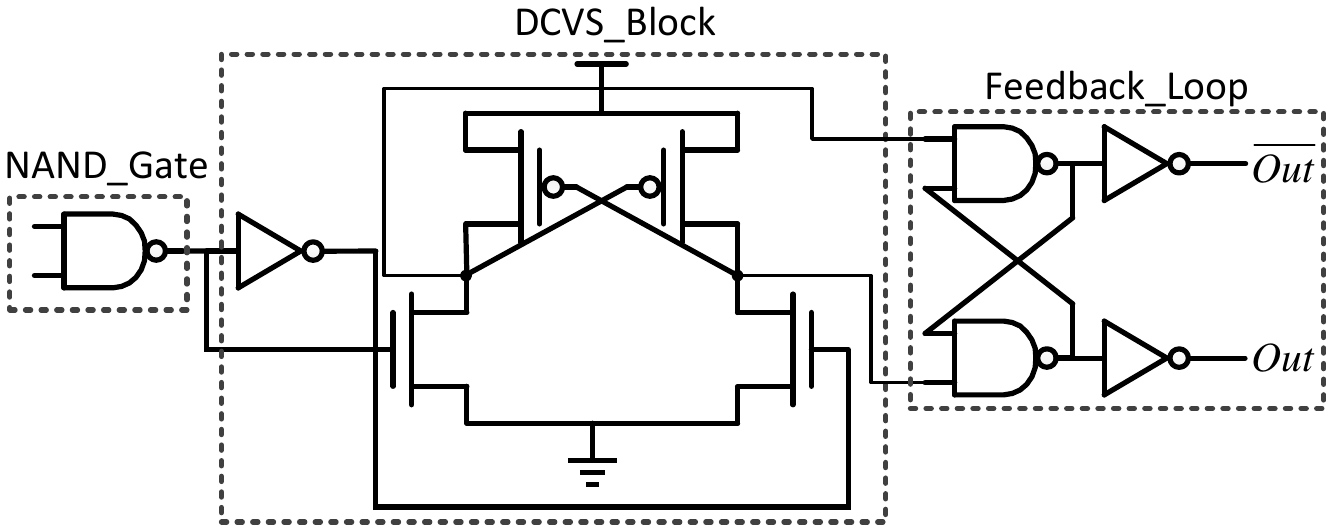}
  \caption{Improved NAND\_gate with MRF and DCVS scheme.}
  \label{fig:fig5_mrf_dcvs}
\end{figure}

\begin{figure}[tp]
  \centering
  \includegraphics[width=3.5in]{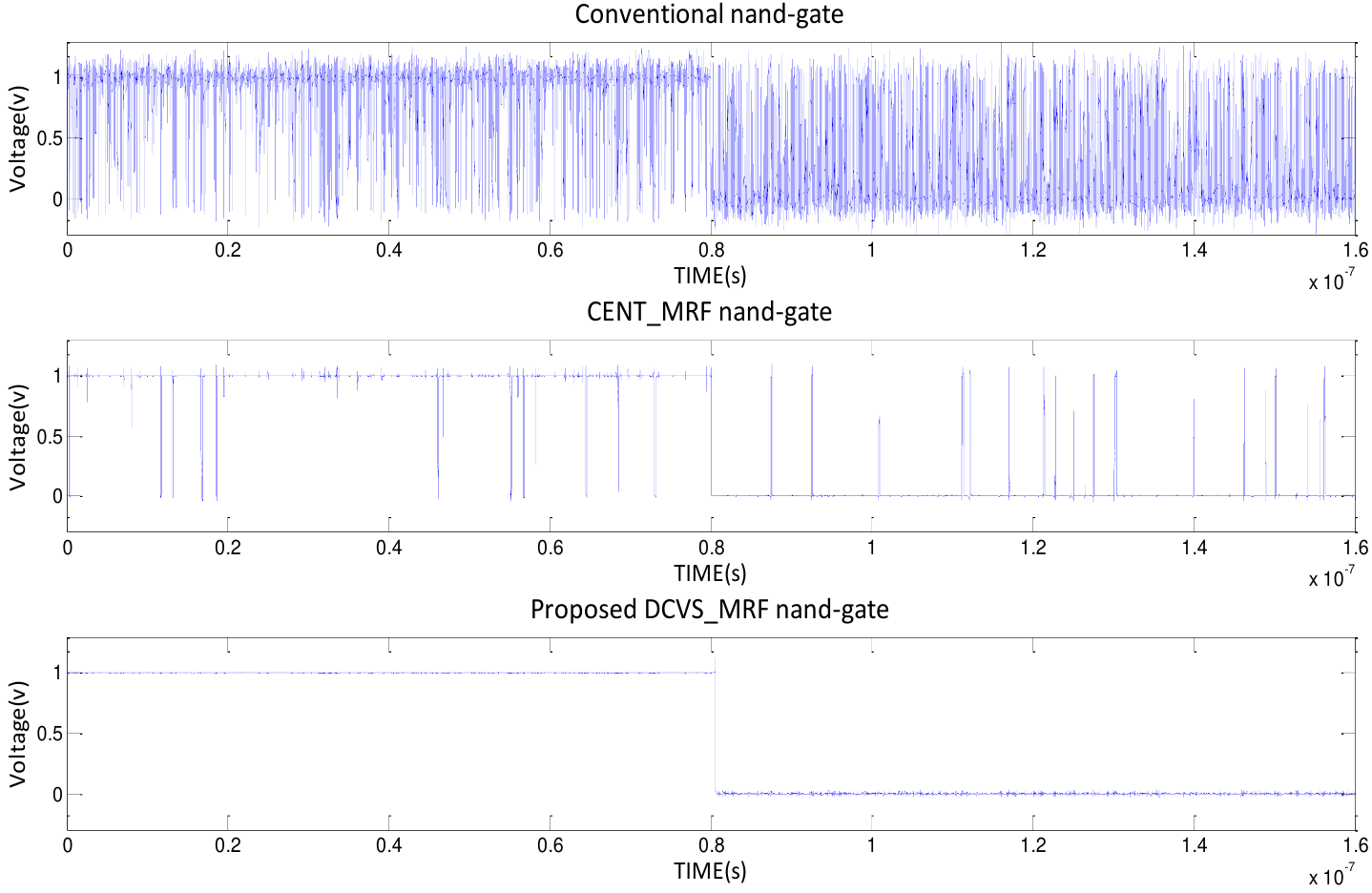}
  \caption{Simulation results for different noise-tolerant circuit scheme.}
  \vspace*{-0.5cm}
  \label{fig:fig6_simulation}
\end{figure}

The DCVS scheme shown in Fig.\ref{fig:fig4_dcvs} also have noise-immunity effect due to its differential operation. Thus, in order to make compensation for the losing immunity in \cite{liu2013general}, we insert the DCVS block into the circuit from Fig.\ref{fig:fig3_liukaikai}(a) which results an improved noise tolerant scheme as shown in Fig.\ref{fig:fig5_mrf_dcvs}. The output of conventional NAND\_Gate is pushed into the DCVS\_Block along with its inverted signal, then the differential output are connected to the Feedback\_Loop which results a mixed circuit scheme based on MRF graph and DCVS technique.

Due to this circuit structure, the effect of noise-immunity will be enhanced significantly as shown in Fig.\ref{fig:fig6_simulation}. Conventional nand-gate, CENT\_MRF nand-gate from \cite{liu2013general} and our proposed DCVS\_MRF nand-gate are simulated with white gaussian noise, in which the SNR of the input signal is $3.5db$. It can be seen that the output of conventional nand-gate is filled with serious disturbance and can hardly be used in logic computation. While the output of CENT\_MRF nand-gate is much better and our proposed one is the best, from which it is well proved that the combination of DCVS and MPF approaches can produce enhanced noise-immunity and provide sufficient compensation to the previous design in \cite{liu2013general}.

\subsection{General Scheme for DCVS\_MRF Logic Circuit}

\begin{figure}[bp]
  \centering
  \vspace*{-0.5cm}
  \includegraphics[width=3.5in]{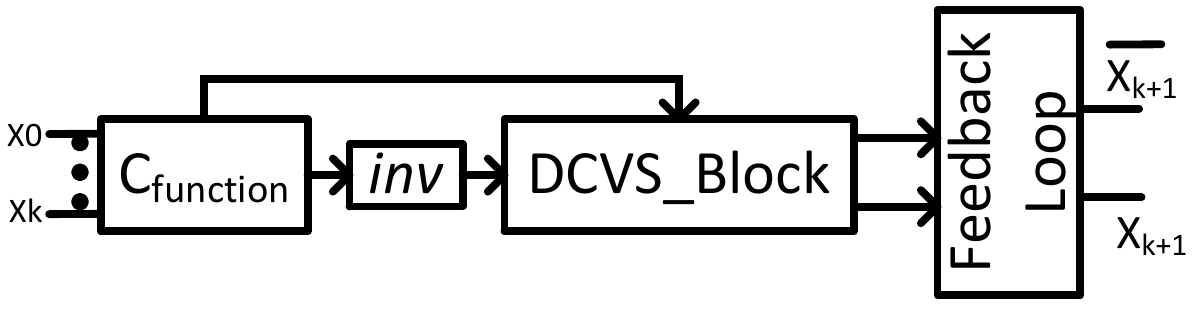}
  \caption{General scheme for proposed DCVS\_MRF circuit design.}
  \label{fig:fig7_general}
\end{figure}

With aforementioned analysis, the general scheme for CENT\_MRF circuit in Fig.\ref{fig:fig3_liukaikai}(b) can be modified by inserting DCVS\_block as shown in Fig.\ref{fig:fig7_general}. The outputs of conventional logic gate $C_{function}$ are inverted by $inv$ and then pushed into the $DCVS\_Block$ with the original input signal. At last, the $FeedbackLoop$ is connected to the $DCVS\_Block$ to construct a complete MRF graph.

For different logic designs, it only needs to replace the $C_{function}$. This methodology is different from \cite{lu2012design} where only the inverter combined MRF and DCVS technique is applied to xor-nxor logic gate design. In our proposed design scheme, the $C_{function}$ in Fig.\ref{fig:fig7_general} can be replaced by any basic logic gates or even some bigger logic blocks. But as pointed out in \cite{nepal2007designing,wey2009design}, the effect of noise-immunity will be weakened if the $C_{function}$ has a big circuit scheme since the $DCVS\_Block$ and $FeedbackLoop$ are not enough to eliminate the influence of the circuit noise.

\section{Simulation and Corresponding results}

\subsection{Introduction of Evaluation Parameter and Simulation Setup}

To evaluate and prove the efficiency of the proposed circuit scheme, Kullback-Leibler distance(KLD) \cite{kullback1997information} is adopted to quantify the noise-immunity of conventional logic gate, CENT\_MRF logic gate in \cite{liu2013general} and our proposed one. In noise-tolerate circuit design, KLD can be used to quantify the difference of two signals based on Eq.\ref{eqution:kld} as described in \cite{liu2013general} where $S_i$ means the output signal without any noise added to its input and $S_r$ means the output signal with a noisy input. $P_{i\_0}$ means the probability of the output as logic "0" for no noise case and other parameters receive their corresponding meanings through its own subscript. From the Eq.\ref{eqution:kld}, it can be seen that the difference of two signals will shrink if the value of KLD is getting smaller, which means that the effect of noise-immunity is stronger with a smaller KLD value.
\begin{equation}
\begin{aligned}
\label{eqution:kld}
KLD(S_i,S_r) = P_{i\_0}log_2(\frac{P_{i\_0}}{P_{r\_0}}) + P_{i\_1}log_2(\frac{P_{i\_1}}{P_{r\_1}})
\end{aligned}
\end{equation}

\subsection{Simulation Results of Noise-Immunity Evaluation}

\begin{figure}[tp]
  \centering
  \includegraphics[width=2in]{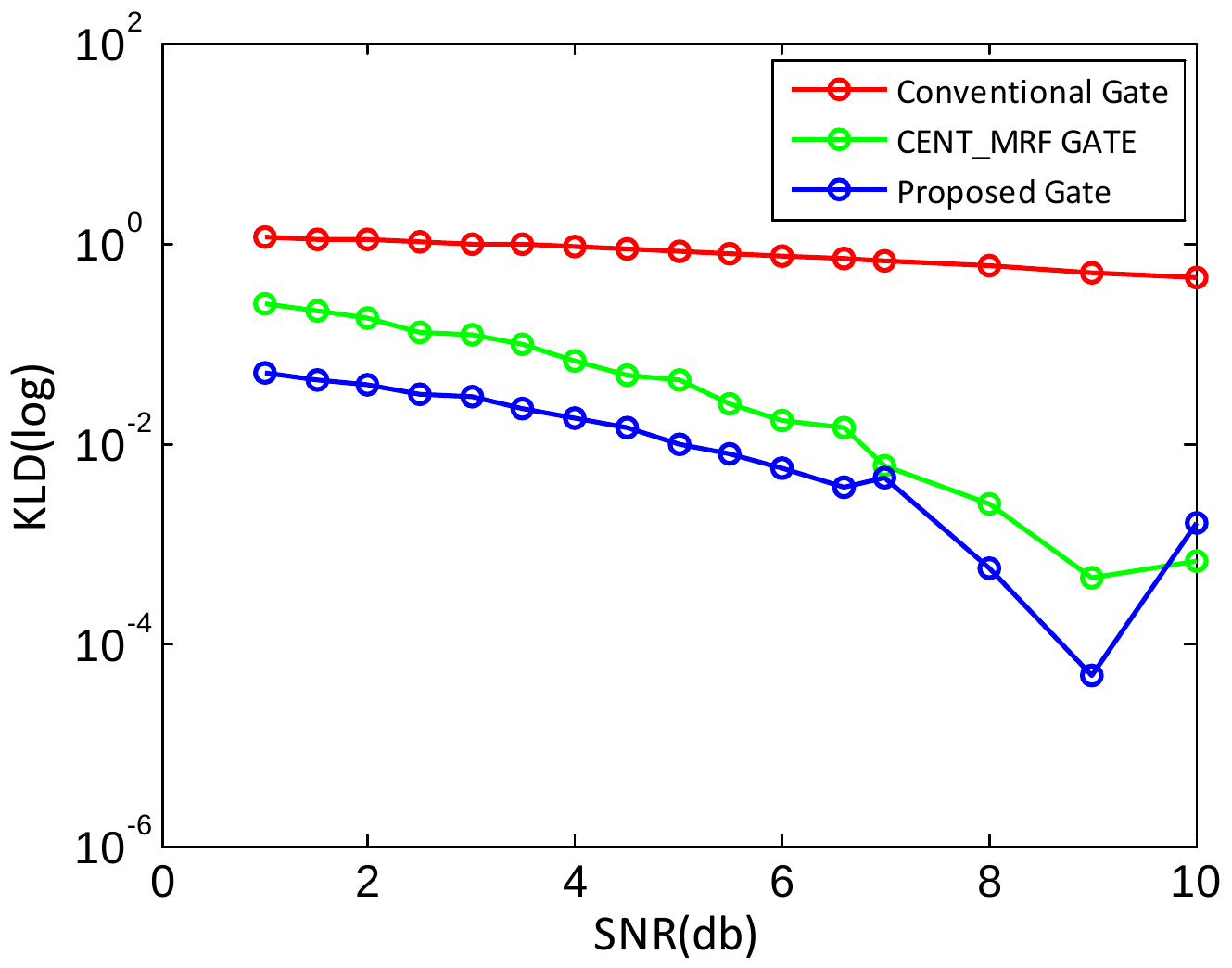}
  \caption{The KLD value of different circuit schemes for $inv$ gate.}
  \vspace*{-0.5cm}
  \label{fig:fig8_inv}
\end{figure}

\begin{figure}[tp]
  \centering
  \vspace*{0.5cm}
  \includegraphics[width=2in]{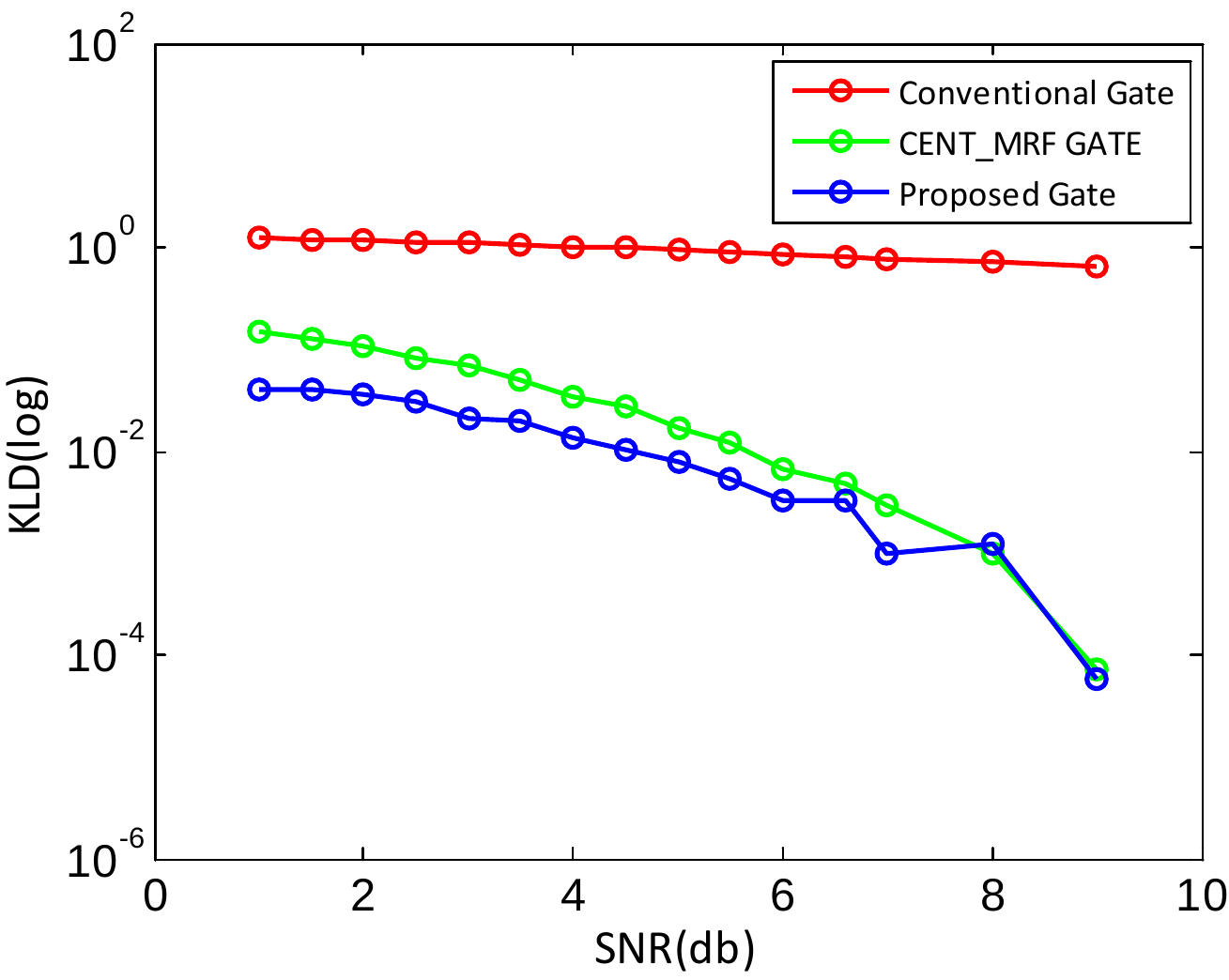}
  \caption{The KLD value of different circuit schemes for $nand$ gate.}
  \vspace*{-0.5cm}
  \label{fig:fig9_nand.}
\end{figure}

\begin{figure}[bp]
  \centering
  \vspace*{-0.5cm}
  \includegraphics[width=2in]{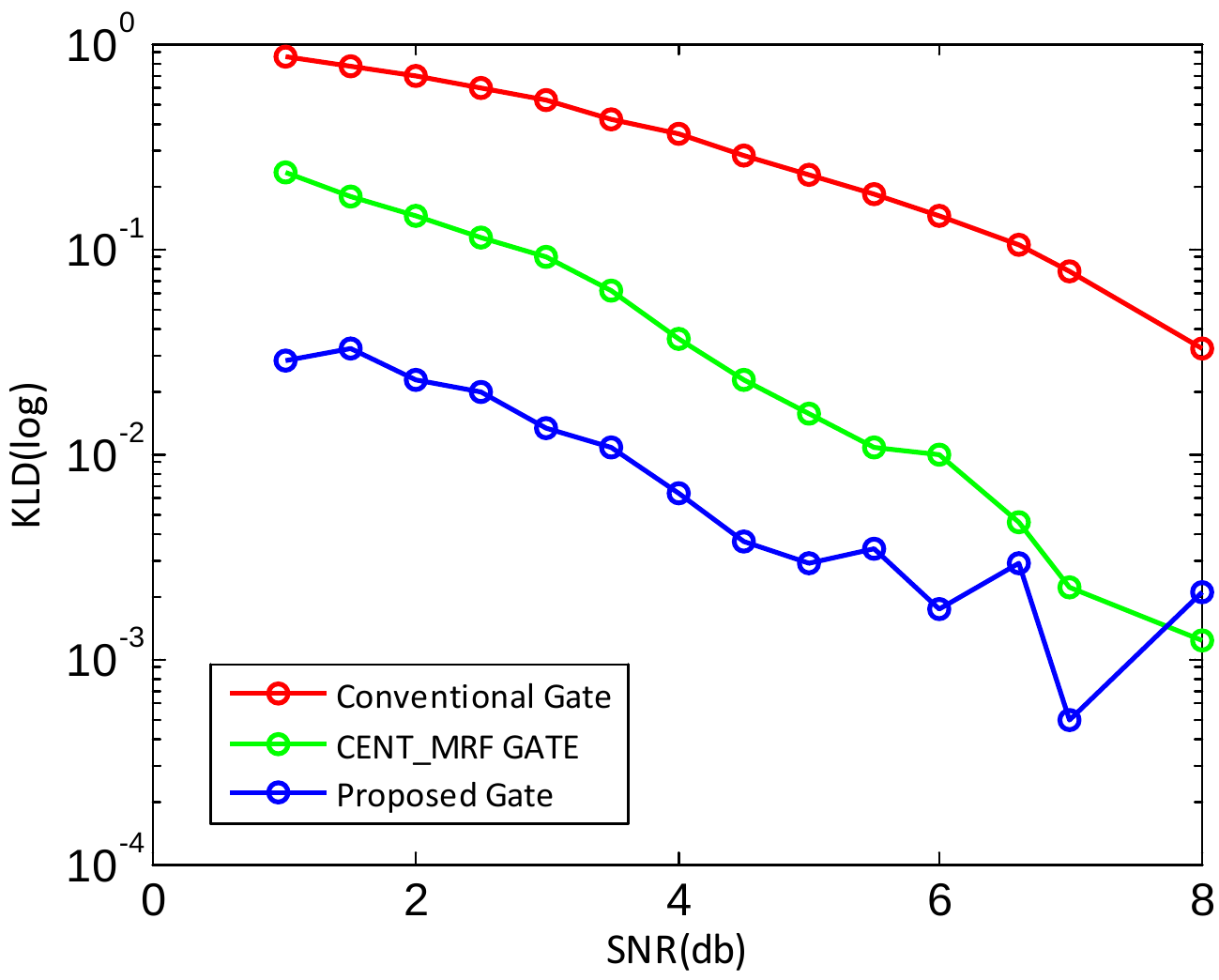}
  \caption{The KLD value of different circuit schemes for $xor$ gate.}
  \label{fig:fig10_xor}
\end{figure}

In order to evaluate the noise-immunity sufficiently, the $inv$, $nand$ and $xor$ gates based on conventional, CENT\_MRF and our proposed methodologies are simulated in Hspice under 65nm technology. The supply voltage is $1V$ and the temperature is $25^{\circ}$C. The input signal are coupled with different levels of white gaussian noise. Then the output signals are pushed into Matlab where the KLD of each simulation will be calculated. The simulation results are shown in Fig.\ref{fig:fig8_inv} to Fig.\ref{fig:fig10_xor}. From the simulation, we find that these proposed logic gates can operate correctly under 1dB SNR which cannot be realized in \cite{liu2013general}. After summing over all the KLD values for each circuit scheme (conventional, CENT\_MRF and Proposed) and computing the average value of KLD, the result shows that the KLD of our design decreases by $76.5\%$ on average than \cite{liu2013general} and superior noise-immunity has been presented.

\subsection{Transistor Number and Power Consumption}

\begin{table}[tp]
  \caption{TRANSISTOR NUMBER AND POWER CONSUMPTION}
  \centering
\begin{tabular}{|c|c|c|}
	\hline
            & \small Transistor\_Num & \small Power ($\mu$W) \\
    \hline
	\small CENT\_MRF\_inv \cite{liu2013general} & 12 & 0.316 \\
    \hline
	\small Proposed\_inv & 16 & 0.318 \\
    \hline
	\small CENT\_MRF\_nand \cite{liu2013general}& 14 & 0.499 \\
    \hline
	\small Proposed\_nand & 18 & 0.497 \\
    \hline
	\small CENT\_MRF\_xor \cite{liu2013general}& 22 & 0.725 \\
    \hline
	\small Proposed\_xor & 26 & 0.721 \\
    \hline
\end{tabular}
  \vspace*{-0.5cm}
  \label{tab:table_transistor_power}
\end{table}

Table \ref{tab:table_transistor_power} shows the transistor number and power consumption of CENT\_MRF gates in \cite{liu2013general} and our proposed ones (inv, nand and xor are listed). Compared to the gates in \cite{liu2013general}, our proposed noise-tolerated gates have few more transistors but the power consumption is quite close, which means that it is very reasonable to sacrifice little performance to get more efficiency of noise-immunity. Thus, our original purpose to make compensation for the losing immunity in \cite{liu2013general} has been successfully achieved with little overhead.

\section{conclusion}

In this paper, a general circuit scheme for noise-tolerant logic design based on MRF theory and Differential Cascode Voltage Switch (DCVS) technique has been proposed. Simulations results show that the KLD value of our design decreases by $76.5\%$ on average than \cite{liu2013general} which means that superior noise-immunity could be obtained with little overhead.

\bibliographystyle{IEEEtran}
\bibliography{mrf-dcvs}

\begin{thebibliography}{10}
\providecommand{\url}[1]{#1}
\csname url@samestyle\endcsname
\providecommand{\newblock}{\relax}
\providecommand{\bibinfo}[2]{#2}
\providecommand{\BIBentrySTDinterwordspacing}{\spaceskip=0pt\relax}
\providecommand{\BIBentryALTinterwordstretchfactor}{4}
\providecommand{\BIBentryALTinterwordspacing}{\spaceskip=\fontdimen2\font plus
\BIBentryALTinterwordstretchfactor\fontdimen3\font minus
  \fontdimen4\font\relax}
\providecommand{\BIBforeignlanguage}[2]{{%
\expandafter\ifx\csname l@#1\endcsname\relax
\typeout{** WARNING: IEEEtran.bst: No hyphenation pattern has been}%
\typeout{** loaded for the language `#1'. Using the pattern for}%
\typeout{** the default language instead.}%
\else
\language=\csname l@#1\endcsname
\fi
#2}}
\providecommand{\BIBdecl}{\relax}
\BIBdecl

\bibitem{nepal2007designing}
K.~Nepal, R.~I. Bahar, J.~Mundy, W.~R. Patterson, and A.~Zaslavsky, ``Designing
  nanoscale logic circuits based on markov random fields,'' \emph{Journal of
  Electronic Testing}, vol.~23, no. 2-3, pp. 255--266, 2007.

\bibitem{wey2009design}
I.-C. Wey, Y.-G. Chen, C.-H. Yu, A.-Y. Wu, and J.~Chen, ``Design and
  implementation of cost-effective probabilistic-based noise-tolerant vlsi
  circuits,'' \emph{Circuits and Systems I: Regular Papers, IEEE Transactions
  on}, vol.~56, no.~11, pp. 2411--2424, 2009.

\bibitem{liu2013general}
K.~Liu, T.~An, H.~Cai, L.~Naviner, J.-F. Naviner, and H.~Petit, ``A general
  cost-effective design structure for probabilistic-based noise-tolerant logic
  functions in nanometer cmos technology,'' in \emph{EUROCON, 2013 IEEE}.\hskip
  1em plus 0.5em minus 0.4em\relax IEEE, 2013, pp. 1829--1836.

\bibitem{lu2012design}
Z.~H. Lu, X.~P. Yu, Y.~Liu, J.~N. Su, and C.~H. Hu, ``Design of nano-scale
  noise tolerant cmos logic circuits based on probabilistic markov random field
  approach,'' \emph{Nanoscience and Nanotechnology Letters}, vol.~4, no.~9, pp.
  914--918, 2012.

\bibitem{kullback1997information}
S.~Kullback, \emph{Information theory and statistics}.\hskip 1em plus 0.5em
  minus 0.4em\relax Courier Dover Publications, 1997.

\bibitem{bhaduri2007reliability}
D.~Bhaduri, S.~K. Shukla, P.~S. Graham, and M.~B. Gokhale, ``Reliability
  analysis of large circuits using scalable techniques and tools,''
  \emph{Circuits and Systems I: Regular Papers, IEEE Transactions on}, vol.~54,
  no.~11, pp. 2447--2460, 2007.

\bibitem{lyons1962use}
R.~E. Lyons and W.~Vanderkulk, ``The use of triple-modular redundancy to
  improve computer reliability,'' \emph{IBM Journal of Research and
  Development}, vol.~6, no.~2, pp. 200--209, 1962.

\bibitem{abraham1974algorithm}
J.~A. Abraham and D.~P. Siewiorek, ``An algorithm for the accurate reliability
  evaluation of triple modular redundancy networks,'' \emph{Computers, IEEE
  Transactions on}, vol. 100, no.~7, pp. 682--692, 1974.

\bibitem{blumer1987occam}
A.~Blumer, A.~Ehrenfeucht, D.~Haussler, and M.~K. Warmuth, ``Occam's razor,''
  \emph{Information processing letters}, vol.~24, no.~6, pp. 377--380, 1987.

\bibitem{gupta2011impact}
V.~Gupta, D.~Mohapatra, S.~P. Park, A.~Raghunathan, and K.~Roy, ``Impact:
  imprecise adders for low-power approximate computing,'' in \emph{Proceedings
  of the 17th IEEE/ACM international symposium on Low-power electronics and
  design}.\hskip 1em plus 0.5em minus 0.4em\relax IEEE Press, 2011, pp.
  409--414.

\bibitem{yang2014iscas}
X.~Yang, F.~Qiao, C.~Liu, Q.~Wei, and H.~Yang, ``Design of multi-stage latency
  adders using detection and sequence-dependence between successive
  calculations,'' in \emph{Circuits and Systems (ISCAS), 2014 IEEE
  International Symposium on}.\hskip 1em plus 0.5em minus 0.4em\relax IEEE,
  2014, pp. 998--1001.

\bibitem{palem2009probabilistic}
K.~V. Palem, P.~Korkmaz, and K.~Kong, ``Probabilistic cmos (pcmos) logic for
  nanoscale circuit design,'' in \emph{International Solid State Circuits
  Conference: Advanced Solid-State Circuits Forum}, 2009.

\bibitem{li1995markov}
S.~Z. Li, \emph{Markov random field modeling in computer vision}.\hskip 1em
  plus 0.5em minus 0.4em\relax Springer-Verlag New York, Inc., 1995.

\end{thebibliography}
\nocite{}
\small

\end{document}